\def\year{2026}\relax
\documentclass[letterpaper]{article} 
\newcommand{{\proj}}[0]{Commenotes}
\usepackage{aaai2026}  
\usepackage{times}  
\usepackage{helvet}  
\usepackage{courier}  
\usepackage[hyphens]{url}  
\usepackage{graphicx} 
\usepackage{natbib}  
\usepackage{caption} 
\usepackage[subrefformat=parens,skip=0pt]{subcaption}
\usepackage[bb=boondox,bbscaled=.95,cal=boondoxo]{mathalfa} 
\usepackage{amsmath} 
\usepackage{framed}
\usepackage{tabularx}
\usepackage{booktabs} 
\DeclareCaptionStyle{ruled}{labelfont=normalfont,labelsep=colon,strut=off} 
\usepackage{algorithm}
\usepackage{algorithmic}

\usepackage{xcolor}

\definecolor{sigboth}{HTML}{2CA02C}   
\definecolor{sigfc}{HTML}{1F77B4}     
\definecolor{signfc}{HTML}{FF7F0E}    
\definecolor{signone}{HTML}{7F7F7F}   

\newcommand{\both}{\textcolor{sigboth}{\textbf{Both}}}
\newcommand{\fconly}{\textcolor{sigfc}{\textbf{FC Only}}}
\newcommand{\nfconly}{\textcolor{signfc}{\textbf{Non-FC Only}}}
\newcommand{\neither}{\textcolor{signone}{Neither}}

\newcommand{\psig}[1]{\textbf{#1}}
\usepackage{newfloat}
\usepackage{listings}
\floatstyle{ruled}
\newfloat{listing}{tb}{lst}{}
\floatname{listing}{Listing}
\title{Fact-Checking Comments Precede Community Notes to Misleading Posts on X}
\author{
    Shuning Zhang\textsuperscript{\rm 1}\thanks{Equal contribution.},
    Dai Shi\textsuperscript{\rm 2}\footnotemark[1],
    Yuwei Chuai\textsuperscript{\rm 3}\thanks{Corresponding authors.},
    Jingruo Chen\textsuperscript{\rm 4},
    Yifan Wang\textsuperscript{\rm 5},\\
    Hui Wang\textsuperscript{\rm 6},
    Gabriele Lenzini\textsuperscript{\rm 3},
    Xin Yi\textsuperscript{\rm 1}\footnotemark[2],
    Hewu Li\textsuperscript{\rm 1}
}
\affiliations{
    \textsuperscript{\rm 1}Tsinghua University, Beijing, China\\
    \textsuperscript{\rm 2}College of Design and Innovation, Tongji University, Shanghai, China\\
    \textsuperscript{\rm 3}SnT, University of Luxembourg, Esch-sur-Alzette, Luxembourg\\
    \textsuperscript{\rm 4}Information Science, Cornell University, Ithaca, New York, United States\\
    \textsuperscript{\rm 5}University of Washington, Seattle, Washington, United States\\
    \textsuperscript{\rm 6}Universität Duisburg-Essen, Essen, Nordrhein-Westfalen, Germany\\
    zsn23@mails.tsinghua.edu.cn
}

\begin{document}


\maketitle


\begin{abstract}
Community-based fact-checking systems, such as $\mathbb{X}$'s Community Notes program, offer a potentially scalable approach against online misinformation diffusion. However, their efficacy is substantially undermined by the delay of fact-check delivery. To address this challenge, we investigated the promise of organic user comments as rapid corrective signals by analyzing a large-scale dataset of over 2.2 million comments directed to 1,841 community fact-checked misleading posts. We developed a high-performance language model pipeline to identify ``fact-checking (FC) comments'', i.e., comments that correct source posts with reasoning or evidence (91\% accuracy). 
Using this pipeline, we find that 99.4\% of misleading posts receive their first FC comments before the official community note is created. Notably, the median time to the initial FC comment is only 0.1 hours, while the creation of community notes has a median latency of 9.7 hours since the publication of misleading posts. Additionally, compared to random posts, those misleading posts with displayed notes have significantly more FC comments. This suggests that FC comments are a unique pattern for misleading posts. We further identified characteristics of FC comments' intensity and speed: (i) FC comments' volume is significantly associated with content richness and misinformation type, favoring multimedia content and missing-context scenarios; (ii) the speed of FC comments remains robust across diverse topics and emotional intensities. These results suggest that platforms could use FC comments as early-warning signals, and potential resource for complementing community notes or automated synthesis.


\end{abstract}

\section{Introduction}

The proliferation of online misinformation poses a critical challenge, with far-reaching negative effects on societies~\cite{xu2023combating,yu2024fake}. Consequently, social media platforms are under persistent pressure to innovate their content moderation strategies for scalable fact-checking. Among emerging approaches, community-based fact-checking has gained increasing attention~\cite{prollochs2022community,drolsbach2024community,chuai2024community}. For instance, $\mathbb{X}$ has introduced the Community Notes program, which allows users to collaboratively add context annotations to potentially misleading posts~\cite{prollochs2022community}. Prior work highlights this system's potential benefits, including increased transparency, the incorporation of diverse perspectives, perceived trustworthiness, and effectiveness in reducing user engagement~\cite{drolsbach2024community,chuai2024community,slaughter2025community}. Despite these strengths, a key limitation of this approach is the lack of timeliness, particularly during the early stage of misinformation dissemination~\cite{chuai2024community}. Community-based fact-checking typically relies on contributors to write notes and reach cross-group consensus, which could cost hours or even days~\cite{guo2022survey}. Consequently, misinformation may spread widely before corrective notes appear, constraining the effectiveness of this community-based approach~\cite{chuai2024did}. 

\begin{figure}[t]
  \centering
  \includegraphics[width=.95\columnwidth]{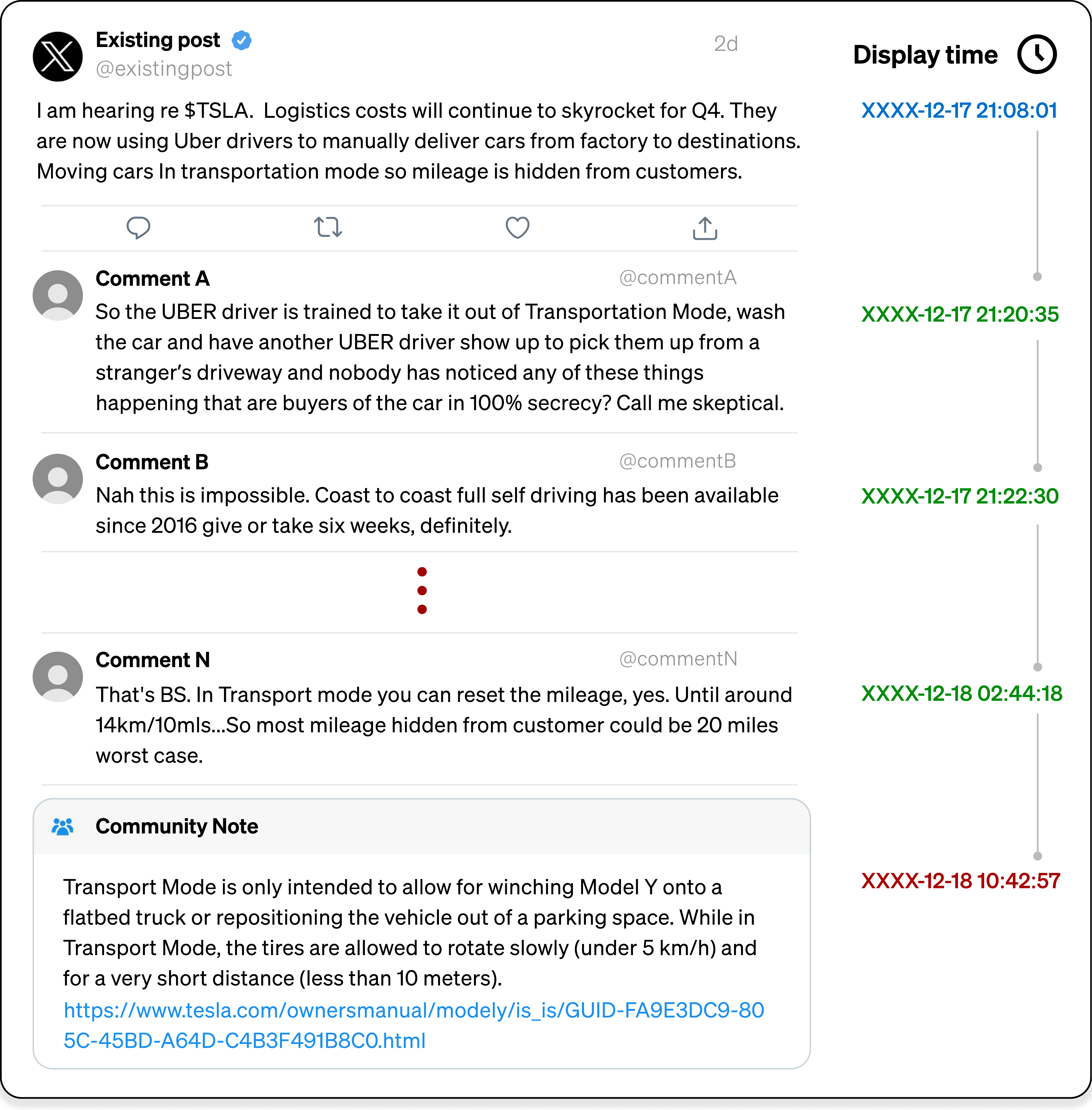}
  \caption{An example of a misleading post followed by its FC comments and displayed community note.}
  \label{fig:header}
\end{figure}

To reduce this latency, we investigate a potentially useful resource~\cite{miyazaki2023fake,jiang2018linguistic}: the real-time fact-checking responses in user comments (Figure~\ref{fig:header}). We hypothesize that user comments, particularly those generated moments after the publication of corresponding posts, may contain early signals of debunking information that precede community notes. Notably, effectively harnessing this data requires a systematic analysis of its existence, timeliness, and characteristics. Therefore, this paper aims to empirically validate whether user comments can serve as a corrective signal for fact-checking. We structure our investigation into three research questions (RQs):

\vspace{.5em}
\noindent \textbf{RQ1 (Feasibility):} \textit{How accurately can language models detect fact-checking intents in comments, and how prevalent are they across misleading posts?}

\noindent \textbf{RQ2 (Timeliness):} \textit{Do fact-checking comments precede community notes, and is this temporal dynamic unique to misleading posts?}

\noindent \textbf{RQ3 (Characteristics):} \textit{What factors are associated with the volume and latency of these fact-checking comments, and how do they differ from general user engagement?}
\vspace{.5em}

To address these RQs, we analyzed $N=$ 2,225,260 direct comments to 1,841 misleading posts, annotated by $\mathbb{X}$'s Community Notes program. We supplemented this dataset with two control groups lacking helpful community notes or containing no notes at all.

Towards RQ1, we proposed an automated pipeline to detect fact-checking (FC) comments, defined as comments that contradict or correct a source post through specific reasoning or external evidence. Specifically, we focus on identifying fact-checking intent over factual veracity to capture early corrective signals, acknowledging that these comments may not always align with professional verification standards. We evaluated multiple Large Language Models (LLMs) and found that Claude-4.5 yielded the best zero-shot performance (93\% accuracy, Cohen's $\kappa$=0.66). To scale this detection, we fine-tuned several language models (e.g., E5, RoBERTa-base) on a LLM-labeled subset of 50,000 comments. The E5 model achieved the highest performance, with 91\% accuracy and a macro average recall of 0.86.

Towards RQ2, our analysis confirms that organic user comments emerge long before official community notes. We found that 99.4\% of misleading posts receive their first FC comment before an official community note is created. The median latency for the first FC comment is just 0.10 hours, compared to 9.7 hours for a community note relative to the post's publication. Robustness checks found 49.8\% of comments with fact-check URLs precede the first community note's creation time. Furthermore, comparative analysis indicates that posts receiving a displayed note have significantly more FC comments than posts with no displayed notes or no community notes. On average, 70 FC comments accumulate before a community note is created, offering potential source for early mitigation.

Towards RQ3, we identified several factors associated with the intensity and speed of FC comments. First, FC comment volume is linked to content richness. While high follower counts increase overall engagement, longer posts ($b = 0.255$) and multimedia content ($b = 0.350$) uniquely associate with FC intensity, where non-FC comments show no such correlation. Second, FC comments' count vary by posts' characteristics. Posts flagged for ``missing context'' uniquely associate with higher counts ($b = 0.177$). Conversely, finance topics ($b = -0.374$), or claims involving satire ($b = -0.302$) and factual errors ($b = -0.188$) significantly suppress FC comments' count. Third, the speed of FC comments is stable across diverse contexts, unlike general engagement, which is sensitive to factors like author verification and media presence.

Collectively, this paper makes three contributions:
(i) We develop a high-performance detection framework using fine-tuned language models to identify FC comments at scale.
(ii) We empirically demonstrate that organic user comments provide rapid corrective potential and emerge earlier than community notes. 
(iii) We provide a characterization of the volume and speed of organic FC comments, identifying post- and comment-level associating factors.


\section{Background and Related Work}

\subsection{User-Centric Evidence in Misinformation Detection}

Current research on using comments for misinformation detection adopts two main strategies: hybrid methods that combine comments with post content, and methods that rely solely on comments. Hybrid approaches often integrate emotional features or textual data from both news posts and their associated comments to improve detection accuracy~\cite{hamed2023fake}. By combining these information sources, such models achieve higher recall than those analyzing news articles in isolation~\cite{albahar2021hybrid,yanagi2020fake}. 

Alternatively, some studies focus exclusively on the comments. For visual misinformation, researchers developed detectors, using the topological, semantic, and linguistic features of comments to distinguish between authentic and manipulated images~\cite{zhang2018fauxbuster,shang2020fauxward}. \citet{jiang2018linguistic} investigated linguistic signals, such as emotional tone and topical patterns, in user comments to detect misinformation. \citet{miyazaki2023fake} characterized spontaneous debunking behaviors towards COVID-19 misinformation. \citet{pilarski2024community} compared the target selection of Community Note program's contributors against users who provide links to professional fact-checks within their comments. Here, we extend the current research scope by quantifying the speed advantage of user comments compared to community notes.

\subsection{Scaling Up Fact-Checking}

Traditional fact-checking by experts is often limited by issues of scalability and public trust~\cite{pennycook2019fighting,straub2022americans}. In response, crowdsourced and automated solutions emerged as scalable alternatives~\cite{kim2020leveraging,quelle2024perils}. For example, initiatives such as the Commuity Notes program invites contributors to write debunking community notes, and uses bridging algorithms to select community notes that are helpful~\cite{chuai2024did}. The Community Notes program has proven effective because it often aligns with expert judgments, resists partisan reasoning, and reaches broad online communities~\cite{allen2021scaling,epstein2020will,micallef2020role}. 

In parallel, automated systems leveraging LLMs have shown potential in supporting various stages of fact-checking, including identifying misleading claims~\cite{qi2024sniffer}, verifying information~\cite{wang2023explainable}, and generating explanations~\cite{yue2024evidence,he2023reinforcement,zeng2024justilm}. While some recent work focuses on synthesizing structured community notes~\cite{de2025supernotes}, our work examines the potential value of organic user comments. Furthermore, LLMs have demonstrated promise in enhancing collective decision-making by summarizing diverse perspectives, mapping opinions to consensus statements~\cite{yang2024llm,burton2024large}, or representing multiple summary viewpoints~\cite{fish2024generative}. However, processing the vast volume of organic social media comments remains a significant challenge. Our paper proposes a framework to extract and utilize organic comments, especially FC comments.

\section{RQ1: Identifying FC Comments}\label{sec:rq1}

To address RQ1, which evaluates the feasibility of detecting fact-checking intent within user comments, we adopted a two-stage method. First, we validated LLMs' capability to identify FC comments by comparing their performance against human annotators. Second, utilizing annotations from the best-performing LLM, we fine-tuned cost-effective models to scale detection across our dataset.

\subsection{Dataset}\label{sec:dataset}

We used the dataset collected by~\citet{chuai2025community}, containing 2,225,260 direct comments across 1,841 source posts. The dataset combines three different sources: community notes, source posts, and user comments. For each post, we acquired the timestamp at which a community note was created, became visible, and the timestamps of all associated comments. While community notes may exhibit status volatility due to rating fluctuations, our analysis focuses on their initial public visibility timestamp. This provides a conservative measurement. By comparing comments against the first instance of a note's display, we ensure that these comments precede subsequent displays. Furthermore, because comments posted after a note becomes visible for the first time may reflect reactions to its corrective information rather than independent fact-checking, including them would confound our analysis. Therefore, we restricted our analysis to comments posted prior to the publication of the corresponding community notes.

\subsection{Task Definition and Annotation Schema}

We operationalized ``FC comments'' as comments that satisfy two conditions: (1) stance, in which the comment clearly challenges, refutes, or questions the validity of the original post, and (2) substance, where the comment provides specific reasoning or evidence rather than mere insults. Different from the multi-class taxonomy used by the Community Notes program, this criteria prioritized the underlying fact-check intent of comments.

We further translated this definition into inclusion and exclusion criteria. Specifically, the inclusion criteria for FC comments required the presence of: (1) external evidence (e.g., links, reports), (2) identification of manipulation (e.g., ``image is photo-shopped''), (3) logical reasoning pointing out contradictions, (4) domain knowledge explaining factual inaccuracies, or (5) corrections of specific facts. Conversely, the exclusion criteria filtered out: (1) agreement with the post (e.g., ``this is true!'', ``good point''), (2) pure insults (e.g., ``you are a liar'', ``OP is dumb''), (3) unsubstantiated slogans (e.g., ``fake news!'') without evidence, (4) general queries (e.g., ``really?''). We implemented these criteria via prompts that required structured JSON output, specifying both the classification label and the refuting reason.

\subsection{Feasibility Validation via LLMs} 

To ensure annotation reliability, we evaluated four state-of-the-art models: gpt-5, claude-4.5, gemini-3, and deepseek-v3.2. We performed reliability checks on a random subset of 1,000 comments. First, we assessed intra-model consistency by prompting each model twice for each comment, where the averaged Cohen's $\kappa$ was 0.88, indicating high consistency. Second, we assessed inter-model consistency, yielding a Fleiss' $\kappa$ of 0.64, indicating high agreement.

To validate the LLM-based approach, we recruited eight trained student assistants to independently annotate a random subsample of 500 comments using the same criteria. The number is determined following prior NLP practices~\cite{zhou2023spatiotemporal,yiliang2025evidenceoutcomes}. They initially annotated 100 comments. Inter-annotator agreement (IAA) among human annotators was moderate ($\kappa$=0.43). They then engaged in a consensus-building session to resolve discrepancies, which primarily focus on distinguishing simple refutations from evidence. During this phase, we provided them the Community Notes program's fact-checking rationales, such as identification of ``outdated information'', as examples. Following this process, they annotated the entire 500 comments. The IAA reached a $\kappa$ of 0.61, indicating substantial agreement.

We then benchmarked multiple LLMs against a human-derived ground truth, constructed by taking the weighted average of human annotations to reflect the aggregate consensus. Each comment was encoded as 0 (non-FC) or 1 (FC), and comments with an average score above 0.5 were treated as FC comment. Under this evaluation scheme, claude-4.5 achieved the highest alignment (Cohen's $\kappa$=0.66, acc=93\%), outperforming gemini-3 ($\kappa$=0.60, acc=89\%), gpt-5 ($\kappa$=0.54, acc=88\%), and deepseek-v3 ($\kappa$=0.50, acc=86\%). Based on this performance, claude-4.5 was selected for performing the final dataset annotation. We randomly sampled another 50,000 comments (with no overlap with human-annotated comments) and annotated them using claude-4.5. This sample size was sufficient for model training to converge, aligning with prior work~\cite{P-stance,conforti-etal-2020-will}. 

\subsection{Scalable Detection}

To enhance computational and financial scalability in identifying FC comments, we fine-tuned several state-of-the-art language models. Our primary classifier was built on E5~\cite{wang2022text}, which was pre-trained on massive text pairs to learn semantic representations. We additionally fine-tuned four models for comparison: BERT~\cite{devlin2019bert}, RoBERTa~\cite{liu2019roberta}, DeBERTa-v3~\cite{he2023debertav}, and SciBERT~\cite{beltagy-etal-2019-scibert}. SciBERT was specifically chosen due to its specialized pre-training on scientific corpora, potentially improving the detection of scientific misinformation in our dataset. We used the human-annotated 100 initial comments described above as the test set.
All models were fine-tuned for 3 epochs with a learning rate of $2\times10^{-5}$, Adam trainer, and a batch size of 16~\cite{devlin2019bert}. Because FC comments constitute a minority class, the dataset exhibits substantial class imbalance, with FC comments comprising 16.1\% (8,050 samples) of the total. To address this, we used a weighted cross-entropy loss function that assigns higher weights to the positive class (FC comments) based on the ratio of negative to positive samples in the training set. This approach forced the models to pay more attention to the minority class, preventing the classifiers from being biased towards the majority class. Training and inference were conducted using an RTX 4090D (24GB GPU), 16 vCPUs, and 80GB of RAM. We evaluated model performance using accuracy, F1-score, Macro recall, and Cohen's kappa. 

\subsection{Results: Model Performance}

Table~\ref{tab:model_comparison} shows the performance of different language models. E5 demonstrated superior performance, achieving an accuracy of 0.912 and a Cohen's $\kappa$ of 0.674. Crucially, it achieved the highest \textit{Macro Avg Recall} of 0.860, indicating a strong capability in recalling potential \textit{FC} and \textit{Non-FC} comments. This recall advantage is particularly important in our setting, where we prioritized minimizing false negatives (i.e., missing FC comments). Among those 100 randomly selected comments, the model identified 18 FC comments, closely aligning with the 15 instances identified by human annotators. Taken together, these results suggest that the fine-tuned language models successfully learned the robust features of FC comments, and were able to identify FC comments without being overwhelmed by the major non-FC class. Consequently, given E5's superior performance and smaller model size, we deployed the E5 model to classify comments.

\begin{table}[!htbp]
    \centering
    \caption{Comparative model performance on the human-annotated 100-comment test set. Values represent the mean $\pm$ standard deviation across five runs with random seeds.}
    \label{tab:model_comparison}
    \small
    \resizebox{\columnwidth}{!}{%
    \begin{tabular}{lcccc}
        \toprule
        \textbf{Model} & \textbf{Kappa ($\kappa$)} & \textbf{F1-score} & \textbf{Accuracy} & \textbf{Recall$^\dagger$} \\
        \midrule
        E5 & \textbf{0.674}$\pm$\textbf{0.031} & \textbf{0.727}$\pm$\textbf{0.025} & \textbf{0.912}$\pm$\textbf{0.011}
        & \textbf{0.860}$\pm$\textbf{0.015} \\
        DeBERTa & 0.655$\pm$0.044 & 0.706$\pm$0.038 & 0.912$\pm$0.011 & 0.828$\pm$0.029 \\
        RoBERTa & 0.645$\pm$0.042 & 0.701$\pm$0.036 & 0.906$\pm$0.011 & 0.835$\pm$0.024 \\
        SciBERT & 0.581$\pm$0.021 & 0.651$\pm$0.016 & 0.882$\pm$0.008 & 0.821$\pm$0.005 \\
        BERT & 0.574$\pm$0.033 & 0.642$\pm$0.027 & 0.884$\pm$0.011 & 0.806$\pm$0.017 \\
        \bottomrule
        \multicolumn{5}{l}{\footnotesize $^\dagger$ Macro Average Recall.} \\
    \end{tabular}
    }%
\end{table}

\section{RQ2: FC Comments Are Faster Than Notes}\label{sec:comments-feasibility} 

In this section, we examine whether user comments can serve as a rapid source of fact-checking evidence, potentially contributing to the Community Notes program.

\subsection{Methodology}\label{sec:methodology}

To analyze the temporal dynamics of community-led corrective comments, we leveraged the fine-tuned E5 model in RQ1 to identify FC comments across the entire dataset. This classifier addresses the class imbalance of FC comments against non-FC comments, achieving high alignment with human judgment ($\kappa$=0.67, Accuracy=0.91). Applying this model, we extracted 336,760 FC comments, counting 15.1\% of all comments.

To quantify temporal patterns, we define three timestamps for each post: post publication, note creation (i.e., the timestamp when a contributor first submits a note), and note display (i.e., the timestamp when a note reaches the consensus threshold for public visibility). We distinguish between creation and display time to capture different stages of the platform's response: creation time is used to measure the actual speed of community note writing. If comments precede this timestamp, they may serve as early evidence or resources for note writers. In contrast, display time is when community note becomes visible to the public. This difference is important as it accounts for the delay caused by the system's consensus mechanisms, during which a note has been written but remains hidden from the general audience.
For Figure~\ref{fig:ts_comments}, we segmented the timeline following post creation into 15-minute intervals, calculating the mean and median volume, and cumulative proportion of FC comments. This allows us to model the emergence and accumulation rate of FC comments provided by users over time.


\subsection{Results}

\begin{figure*}[!htbp]
    \centering
    \begin{subfigure}{0.49\textwidth}
    
    \includegraphics[width=\textwidth]{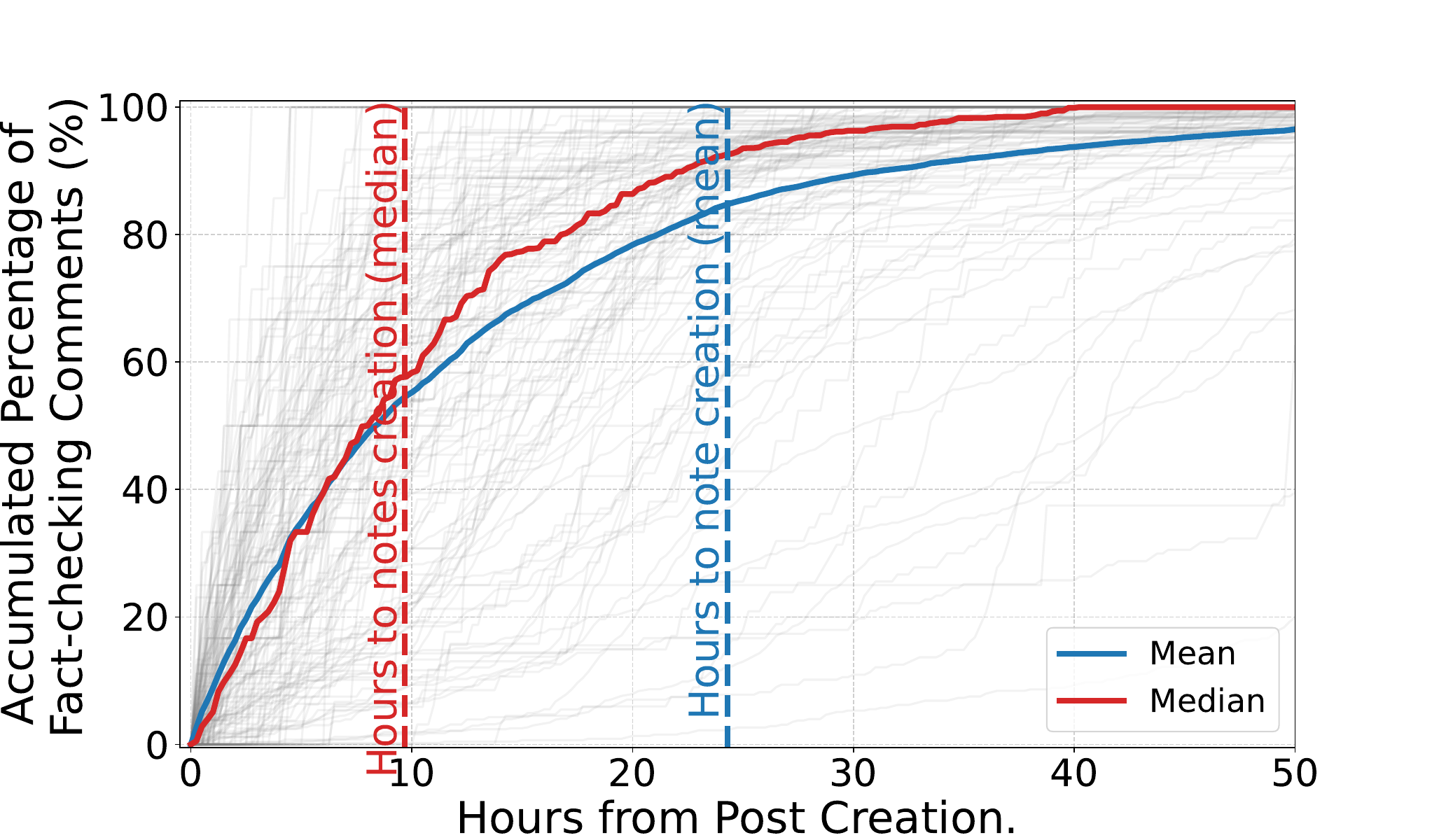}
    \caption{}
    \label{fig:images-all-percentage}
    \end{subfigure}
    \hfill
    \begin{subfigure}{0.49\textwidth}
    
    \includegraphics[width=\textwidth]{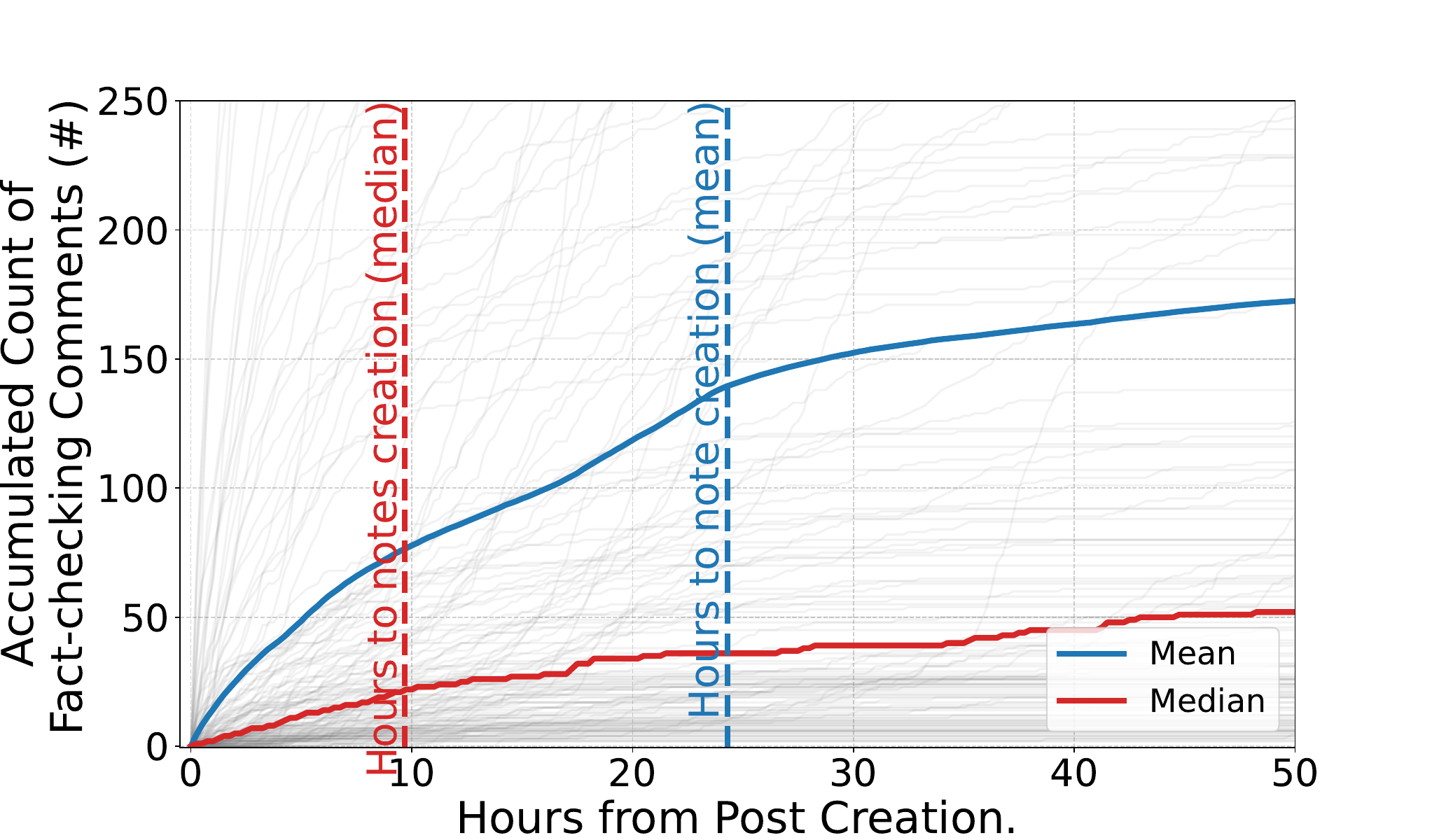}
    \caption{}
    \label{fig:images-all-count}
    \end{subfigure}
    \caption{FC comments' number and their proportion relative to the cumulative total of FC comments observed up to the collection of the dataset. (a)~Cumulative percentage of FC comments. (b)~Cumulative count of FC comments over time.}
    \label{fig:ts_comments}
\end{figure*}

Figure~\ref{fig:ts_comments} shows that a substantial volume of FC comments appears early in a post's lifecycle, often long before an official community note is created. As illustrated in Figure~\ref{fig:images-all-percentage}, FC activity appears rapidly, with over 99\% of misleading posts receiving their first FC comment before the first community note is created (99.4\%) or displayed (99.5\%). 99.3\% of posts receive their FC comment within two hours of publication. With a median latency of 0.10 hours, the cumulative distribution of FC comments exceeds 50\% significantly ahead of the 9.7-hour median creation time for community notes. 

Corroborating this trend, Figure~\ref{fig:images-all-count} shows the number of FC comments that accumulate before a note's creation. At the median note creation time, the mean cumulative count of these comments is approximately 70, with a median of 15.5. Furthermore, the accumulation of FC comments is front-loaded rather than linear. The concave shape of the cumulative count (Figure~\ref{fig:images-all-count}) shows that the increasing rate of FC comments is highest immediately following the post's creation. These findings suggest that substantial FC comments emerge early in the misinformation lifecycle, offering potential for automated synthesis.

\textbf{Examining comments with fact-check URLs:}
To provide additional evidence, we identified FC comments that embed verification URLs directly within their text. A URL is classified as a fact-check URL if its domain includes the string ``fact-check'' or ``factcheck'', or if it appears in Wikipedia's representative fact-checking website lists\footnote{https://en.wikipedia.org/wiki/List\_of\_fact-checking\_websites}. Applying this criteria, we extracted 1,740 valid comment-URL pairs, yielding 1,734 unique comments across 359 distinct posts. This preliminary identification relies on explicit URL patterns, which likely represents a conservative lower bound, where many FC comments may contain verification or reference external sources without direct links.

Temporal analysis of these comments reveal patterns aligning with our findings on fact-checking activity. Compared against the creation time of the first note, 867 comments (49.8\%) preceded creation, and 873 (50.2\%) were posted at or after this timestamp. Relative to the display time of the first community note, 1,384 comments (79.5\%) were posted beforehand and 356 (20.5\%) appeared concurrently or subsequently. Collectively, these results show the FC comments' potentials for augmenting official community notes.

\textbf{Comparison with random comments:}
To explore the generalizability of our findings, we conducted a preliminary analysis using two control groups: (1) non-community-note posts, which received neither note requests nor community notes, (2) no-displayed-note posts, which contained posts that failed to achieve ``Helpful'' status for public display. To mitigate temporal confounding, we used $\mathbb{X}$ search API to sample data from Apr 2023 and Apr 2026. For each period, we sampled five posts per day across ten randomly selected days. We used the detector developed in RQ1 to identify FC comments within these datasets. The non-community-note group yielded 100 posts (95,024 comments, 1,382 FC comments), while the no-displayed-note group yielded 100 posts (98,675 replies, 19,815 FC comments).

\begin{figure*}[!htbp]
    
    \includegraphics[width=1.0\textwidth]{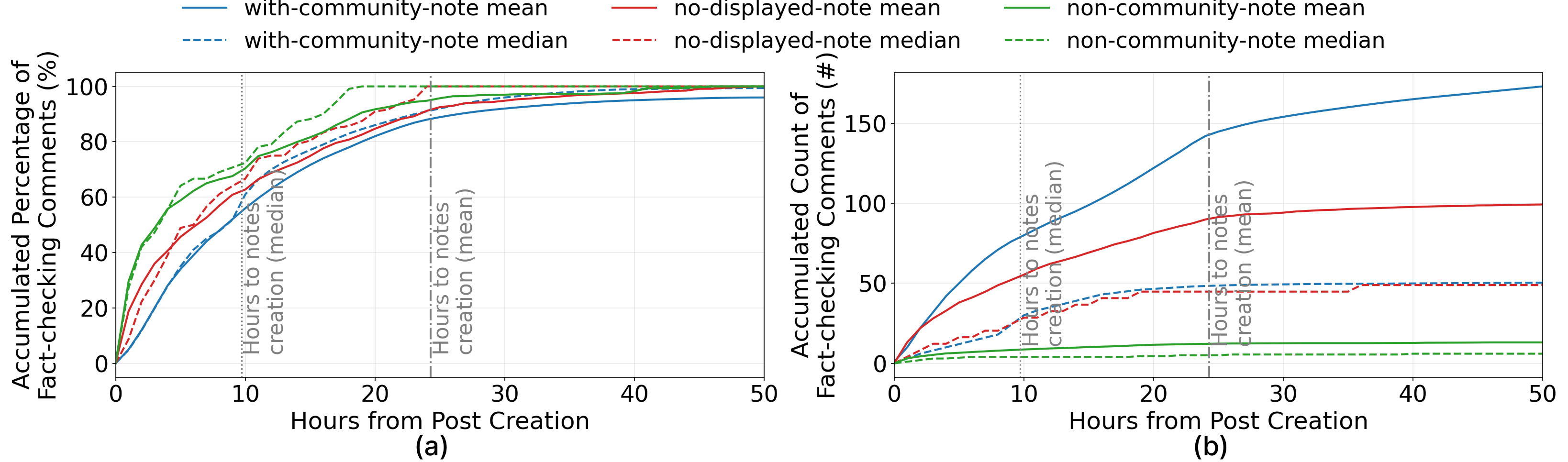}
    \caption{(a) The distributions of accumulated percentages and (b) accumulated counts for FC comments over time. The distributions of FC comments to posts with (i) community notes written and displayed, (ii) community notes written but no one displayed, and (iii) no community notes written or requested are shown, respectively.}
    \label{fig:comparison}
\end{figure*}

Figure~\ref{fig:comparison} shows that while FC comments rapidly emerge across all three groups, their magnitudes and accumulation dynamics diverge significantly. Within the first two hours, 99.3\% of posts with community notes received at least one FC comment, significantly higher than the 88.10\% for the no-displayed-note group and 79.38\% for the non-community-note group (two-proportion z-tests, $p < .001$, Holm-corrected).
Furthermore, at the first community note's median creation time, posts with displayed notes accumulated a mean of 70 FC comments (median=15.5), substantially exceeding both the no-displayed-note (mean=13.35, median=6) and non-community-note groups (mean=8.60, median=4).
This gap also persists at the first community note's mean display time, compared with the no-displayed-note group (mean=22.24, median=11), and non-community-note group (mean=12.19, median=5).
Mann-Whitney U tests confirm significant differences across all pairs ($p < .001$, Holm-corrected). Collectively, these comparison suggests that FC comments are uniquely tied to community notes rather than a pervasive phenomenon.

\section{RQ3: FC Comments' Intensity and Efficiency}

\subsection{Methodology}

We analyze two dimensions of comments: \textbf{intensity}, defined as the total number of comments a post receives, and \textbf{efficiency}, defined as the time lag between the earliest comment and the creation of the first community note. Our analysis follows a two-step approach. First, we conduct a descriptive analysis to characterize the FC comments' intensity and efficiency across topics and post popularity. Here, we use Spearman's rank correlation ($\rho$) for popularity-based associations, Welch's t-test for account verification status, and the Kruskal--Wallis test for multi-topic comparisons. Second, we use fixed-effects regression to identify post-, community-note-, and comment-level factors that correlate with comments' intensity and efficiency. Community-note-level factors include attributes such as the community note's length and helpfulness ratings. To isolate general engagement effects, we compare FC comments' findings against those of non-FC comments and all comments.



We categorized the features into four levels to capture the multifaceted nature of comments. For the descriptive analysis, we operationalized post visibility using a popularity proxy, $\log_2(c/h)+1$, where $c$ is the number of comments and $h$ is the hours since post creation~\cite{wu2017sequential}.

\textbf{Content and topical context:} We measured \textit{Word Count}, \textit{Media} presence, and \textit{Topic} (Politics, SciTech, Entertainment, and Finance) to account for the complexity and domain-specific interest of the post~\cite{chuai2024did}. 

\textbf{Author authority and experience:} We used \textit{Follower/Following Count}, \textit{Account Age} and \textit{Verification Status} as proxies for the source author's platform influence. For note writers, we include \textit{Author\_Prev\_Count} to reflect their prior fact-checking experience, similar to~\citet{chuai2024did}. 

\begin{figure*}[!htbp]
    \centering
    \includegraphics[width=0.9\textwidth]{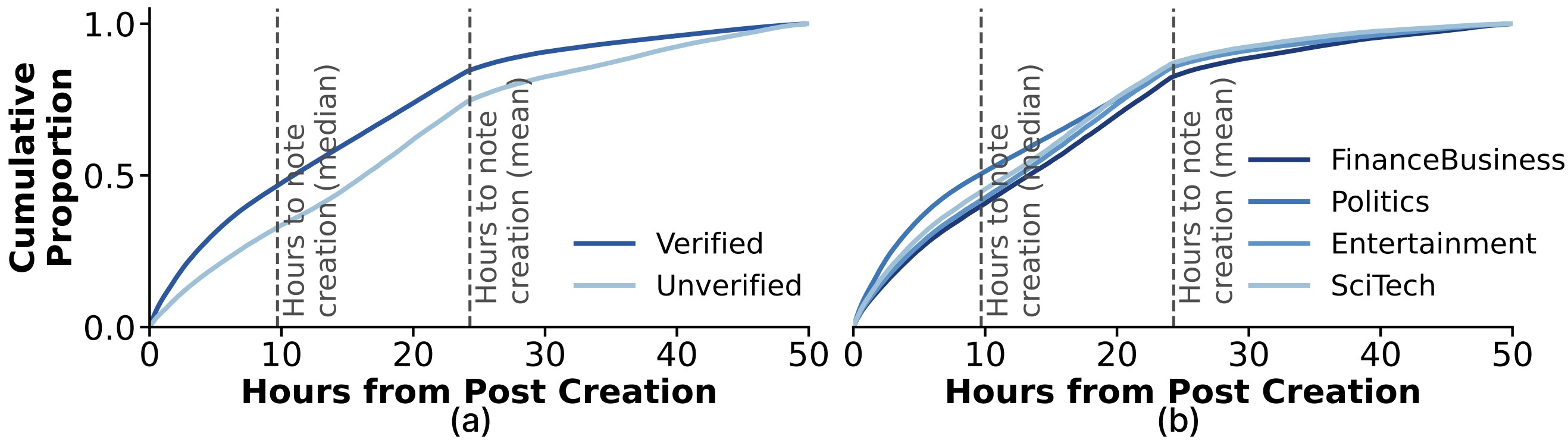}
    
    \caption{Cumulative proportion distributions of FC comments, categorized by the post's (a) author account verification status and (b) topic.}
    \label{fig:deep}
\end{figure*}

\textbf{Affective dimensions:} Using a pre-trained RoBERTa model~\cite{chuai2025community,chuai2024community}, we extracted the intensity of four discrete emotions (\textit{Anger}, \textit{Joy}, \textit{Fear}, \textit{Sadness}) from both source posts and comments to investigate how emotional signaling influences crowd-sourced fact-checking. 

\textbf{Community-note-specific dimensions:} We included binary indicators of the community note's rationale (e.g., \textit{Factual Error}, \textit{Missing Context}) and quality metrics (\textit{Is\_Helpful}, \textit{Note\_Length}) to assess how the nature of misinformation correlates with the comments.

All count-based metrics (e.g., volume, followers) were log-transformed to handle skewed distributions and ensure statistical robustness. 

\textbf{Modeling fact-check intensity:} To analyze how post- and community-note-level factors influence the volume of comments (e.g., FC and non-FC comments) prior to note creation, we employed a fixed-effects regression model. This approach allows us to control for unobserved author-specific traits that might otherwise bias results. The model is specified as follows:

\begin{equation}\tag*{}
\label{eq:model_count}
\begin{aligned}
    &\log(\text{Count}_i) = \alpha_{\text{Note\_Author}} \\
    & + \underbrace{\beta_1 \text{Followers}_{\text{src}} + \beta_2 \text{Verified}_{\text{src}} + \beta_3 \text{Words}_{\text{src}} + \beta_4 \text{Media}}_{\text{Post Traits}} \\
    & + \underbrace{\beta_5 \text{Anger}_{\text{src}} + \beta_6 \text{Joy}_{\text{src}} + \beta_7 \text{Fear}_{\text{src}} + \beta_8 \text{Sadness}_{\text{src}}}_{\text{Post Emotions}} \\
    & + \underbrace{\sum_{t \in \text{Topics}} \beta_{t}\text{Topic}_{i,t}}_{\text{Topics}} \\
\end{aligned}
\end{equation}

\begin{equation}
\label{eq:model_count}
\begin{aligned}
    & + \underbrace{\gamma_1 \text{Note\_Length}_{ij} + \gamma_2 \text{Has\_TrustWorthy\_Sources}_{ij}}_{\text{Community Note Traits (Attributes)}} \\
    & + \underbrace{\gamma_3 \text{Author\_Prev\_Count}_{ij} + \gamma_4 \text{Is\_Helpful}_{ij}}_{\text{Community Note Traits (Quality)}} \\
    & + \underbrace{\gamma_5 \text{Reason\_Factual\_Error}_{ij} + \gamma_6 \text{Reason\_Missing\_Context}_{ij}}_{\text{Community Note Traits (Reasons I)}} \\
    & + \underbrace{\gamma_7 \text{Reason\_Unverified\_Claim}_{ij} + \gamma_8 \text{Reason\_Satire}_{ij}}_{\text{Community Note Traits (Reasons II)}} + \epsilon_{ij}
\end{aligned}
\end{equation}

In Equ~\ref{eq:model_count}, the indices $i$ and $j$ denote the $i$-th post and the $j$-th community note, respectively. The dependent variable $\log(\text{Count}_i)$ is the log-transformed count of comments (e.g., FC comments, non-FC comments) received by post $i$ before the note was created. $\alpha_{\text{Note\_Author}}$ captures the fixed effects for note authors.

For post-level features, the subscript $src$ denotes attributes of the source post or its author. Specifically, $\text{Followers}_{\text{src}}$ represents the log-transformed number of followers of the post author, and $\text{Verified}_{\text{src}}$ is a binary indicator for verification status. $\text{Words}_{\text{src}}$ denotes the log-transformed word count of the original post, while $\text{Media}$ indicates the presence of multimedia. We also include emotional scores ($\text{Anger}_{\text{src}}$, etc.) and topic indicators ($\text{Topic}_{i,t}$) for the source post, following~\citet{chuai2024did,chuai2024community}. 

At community note level, $\text{Note\_Length}_{ij}$ is the log-transformed length of community note $j$. $\text{Has\_Trustworthy\_Sources}_{ij}$ indicates whether the note cites credible sources. $\text{Author\_Prev\_Count}_{ij}$ is the log-transformed number of prior community notes written by the author, representing writer experience. $\text{Is\_Helpful}_{ij}$ is a binary outcome indicating if the note was rated helpful. The $\text{Reason}$ variables (e.g., $\text{Reason\_Factual\_Error}_{ij}$) are binary variables for the specific reason flagged by the community note. 

\textbf{Modeling fact-checking delay:} To examine the factors associated with comments' speed relative to the community note, we employed a multiple linear regression model. The dependent variable, $\text{log}(\text{Delay}_{ik})$, represents the time between the first comment, and the creation of the earliest note. Note that for FC comment, non-FC comment or all comment, we used the same model.

\begin{equation}
\label{eq:model_delay}
\begin{aligned}
    & \log(\text{Delay}_{ik}) = \alpha \\
    & + \underbrace{\beta_1 \text{Followers}_{\text{src}} + \beta_2 \text{Verified}_{\text{src}} + \beta_3 \text{Words}_{\text{src}} + \beta_4 \text{Media}}_{\text{Post Traits}} \\
    & + \underbrace{\beta_5 \text{Anger}_{\text{src}} + \beta_6 \text{Joy}_{\text{src}} + \beta_7 \text{Fear}_{\text{src}} + \beta_8 \text{Sadness}_{\text{src}}}_{\text{Post Emotions}} \\
    & + \underbrace{\sum_{t \in \text{Topics}} \beta_{t}\text{Topic}_{i,t}}_{\text{Topics}} \\
    & + \underbrace{\delta_1 \text{Words}_{\text{comment}} + \delta_2 \text{Anger}_{\text{comment}} + \delta_3 \text{Joy}_{\text{comment}}}_{\text{Reply Traits (Content)}} \\
    & + \underbrace{\delta_4 \text{Fear}_{\text{comment}} + \delta_5 \text{Sadness}_{\text{comment}}}_{\text{Reply Traits (Emotions)}} \\
    & + \underbrace{\gamma_1 \text{Note\_Length}_{i} + \gamma_2 \text{Has\_TrustWorthy\_Sources}_{i}}_{\text{Community Note Traits (Attributes)}} \\
    & + \underbrace{\gamma_3 \text{Reason\_Factual\_Error}_{i} + \gamma_4 \text{Reason\_Missing\_Context}_{i}}_{\text{Community Note Traits (Reasons I)}} \\
    & + \underbrace{\gamma_5 \text{Reason\_Unverified\_Claim}_{i} + \gamma_6 \text{Reason\_Satire}_{i}}_{\text{Community Note Traits (Reasons II)}} + \epsilon_{ik} 
\end{aligned}
\end{equation}

In Equ~\ref{eq:model_delay}, the indices $i$ and $k$ denote the $i$-th post and the $k$-th comment, respectively. The subscript $src$ refers to attributes of the source post. $\text{Followers}_{\text{src}}$ represents the log-transformed follower count of the post author, and $\text{Verified}_{\text{src}}$ indicates their verification status. $\text{Words}_{\text{src}}$ is the log-transformed word count of the post, while $\text{Media}$ indicates the presence of multimedia. We also include emotional scores ($\text{Anger}_{\text{src}}$, etc.) and topic indicators ($\text{Topic}_{i,t}$) for the source post. 

At the comment level, $\text{Words}_{\text{comment}}$ denotes the log-transformed word count of the comment $k$. $\text{Anger}_{\text{comment}}$, $\text{Joy}_{\text{comment}}$, $\text{Fear}_{\text{comment}}$, and $\text{Sadness}_{\text{comment}}$ represent the emotions expressed in the comment. 

At community note level, $\text{Note\_Length}_{i}$ represents the log-transformed length of the fist community note. $\text{Has\_Trustworthy\_Sources}_{i}$ indicates if the note cites credibility resources. The $\text{Reason}$ variables (e.g., $\text{Reason\_Factual\_Error}_{i}$) act as controls for the type of misinformation identified, which may affect the complexity of the verification process.

\subsection{Results}

\textbf{Distribution of FC comments by topics:}
For clarity, we define the metrics used in this analysis as follows: \textit{Count} refers to the absolute number of FC comments on a post, measured at the time of the first community note's creation. \textit{Proportion} refers to the ratio of these FC comments to the total number of comments on the post at that same time point.

\begin{figure}[!htbp]
    \centering
    \includegraphics[width=0.45\textwidth]{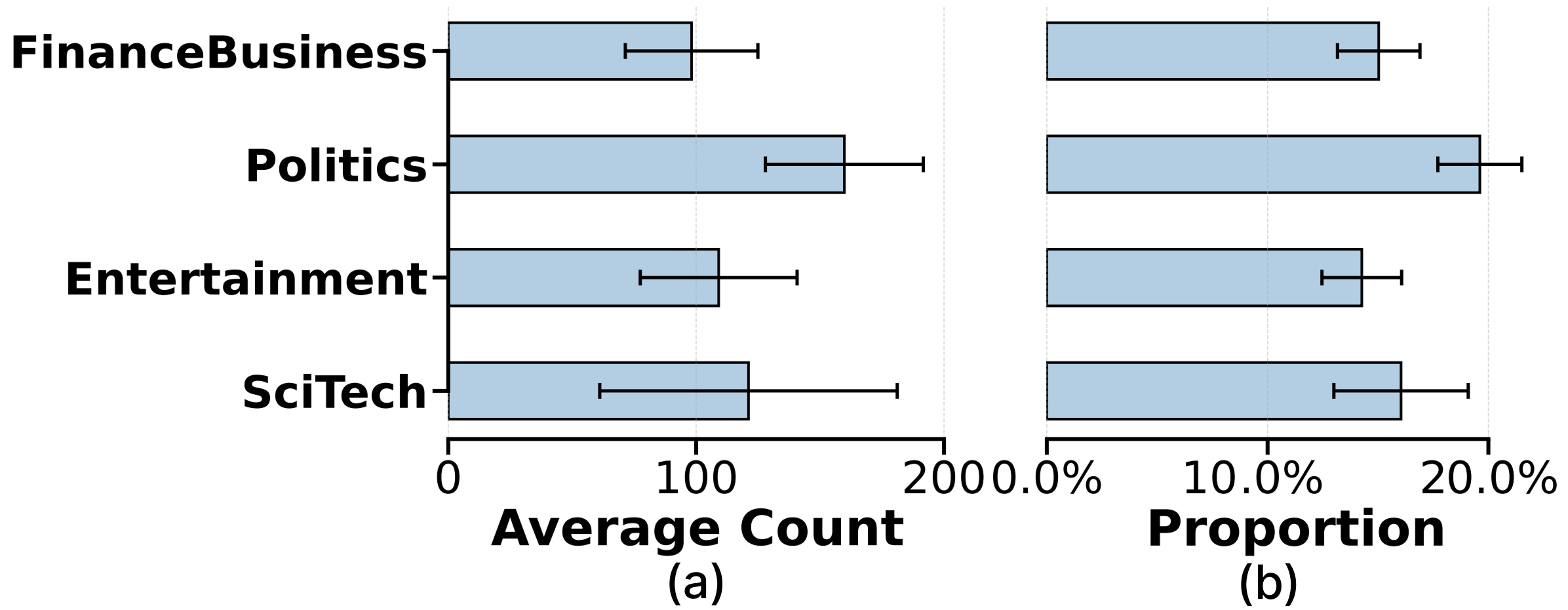}
    
    \caption{
    (a) Mean count and (b) proportion (relative to all comments) of FC comments by topic at note creation time. Error bars denote one standard deviation.}
    \label{fig:topic_comparison}
\end{figure}

\begin{figure*}[!htbp]
    \centering 
    
    \includegraphics[width=\textwidth]{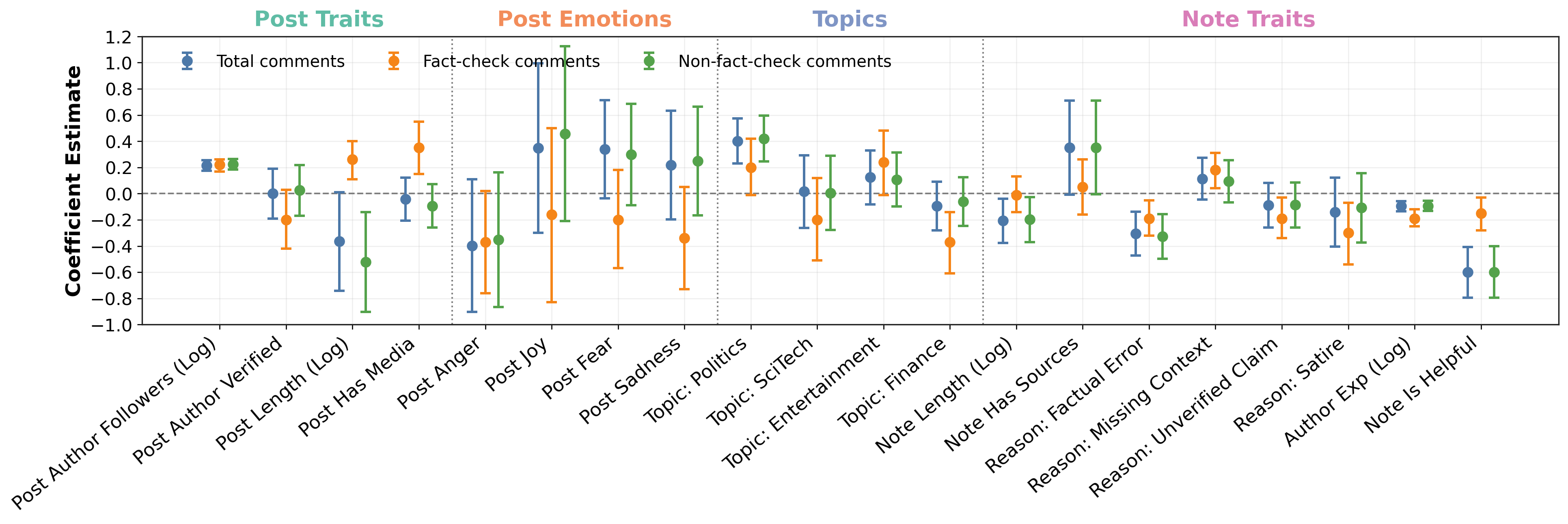}
    \caption{Regression coefficients for FC, all and non-FC comment count (i.e., intensity) across post-, comment-, and community-note-level factors. Error bars indicate 95\% Confidence Intervals (CIs).}
    \label{fig:debunking_count}
\end{figure*}

We first examined the distribution of posts across four primary topics. The dataset was primarily composed of \textit{Finance \& Business} (30.49\%), followed by \textit{Politics} (28.30\%), \textit{Entertainment} (23.53\%), and \textit{Science \& Technology} (10.53\%). Note that each post could belong to multiple categories. As shown in Figure~\ref{fig:topic_comparison}, although posts related to \textit{Politics} attracted the highest  FC comments' count, a Kruskal-Wallis test demonstrated no significant difference for FC comments' count across topics ($H=4.250$, $p=.236$). Similarly, the proportion of FC comments did not vary significantly among categories ($H=1.651$, $p=.648$). 

Furthermore, the cumulative distribution of comments (Figure~\ref{fig:deep}) shows a high degree of overlap across all four topics. This temporal consistency suggests that the speed at which FC comments accumulate is independent of the subject topics. Collectively, these findings indicate that FC comments exist robustly, distributed across diverse domains, rather than concentrated in specific controversial subjects.

\textbf{Visibility and engagement effects:}
Post popularity was strongly correlated with the absolute number of FC comments received before community note creation ($\rho = 0.8925$, $p < .001$), but showed a weak negative correlation with their proportion relative to total comments at the time of note creation ($\rho = -0.1346$, $p < .001$).

Similarly, we examined the effects of author verification status. A Welch's t-test confirmed that posts by verified authors received a higher mean number of FC comments before note creation (M=90.36) than posts by unverified authors (M=65.83), but this difference was not statistically significant ($t = 1.292$, $p = .197$). In contrast, the proportion of FC comments was significantly lower for posts with verified authors (16.28\%) than for posts with unverified authors (19.41\%, $t = -4.152$, $p < .001$). This suggests that before note creation, while verified accounts attract more comment engagement, FC comments make up a smaller share.

Crucially, the temporal analysis in Figure~\ref{fig:deep} reveals that the accumulation curve for verified authors rises more sharply than that of unverified authors. This indicates that FC comments on those posts with verified profiles have higher volumes and accumulate more rapidly, reflecting an immediate community response.

\textbf{Analysis of FC comments' intensity:}
As seen in Figure~\ref{fig:debunking_count}, FC comment volume exhibits unique correlations with specific post characteristics, such as content length and multimedia presence. Longer posts ($b=0.255$, $p <.001$) and those containing media ($b=0.350$, $p <.001$) uniquely attract more FC comments. Conversely, non-FC comments show no significant correlation with either post length ($b=0.062$, $p = .36$) or media presence ($b=-0.051$, $p=.55$). These artifacts trigger more corrective responses  perhaps as they provide more material for users to challenge. In terms of general engagement, an author's follower count positively correlates with both FC ($b=0.212$, $p < .001$) and non-FC comments ($b=0.228$, $p < .001$), while author verification status lacks a significant relationship with either FC ($b=-0.199$, $p = .08$) or non-FC comments ($b=0.004$, $p = .97$). These shared patterns suggest that, high visibility broadly increases overall comment volume rather than uniquely influences FC comments.


\begin{figure*}[!htbp]
    
    \includegraphics[width=\textwidth]{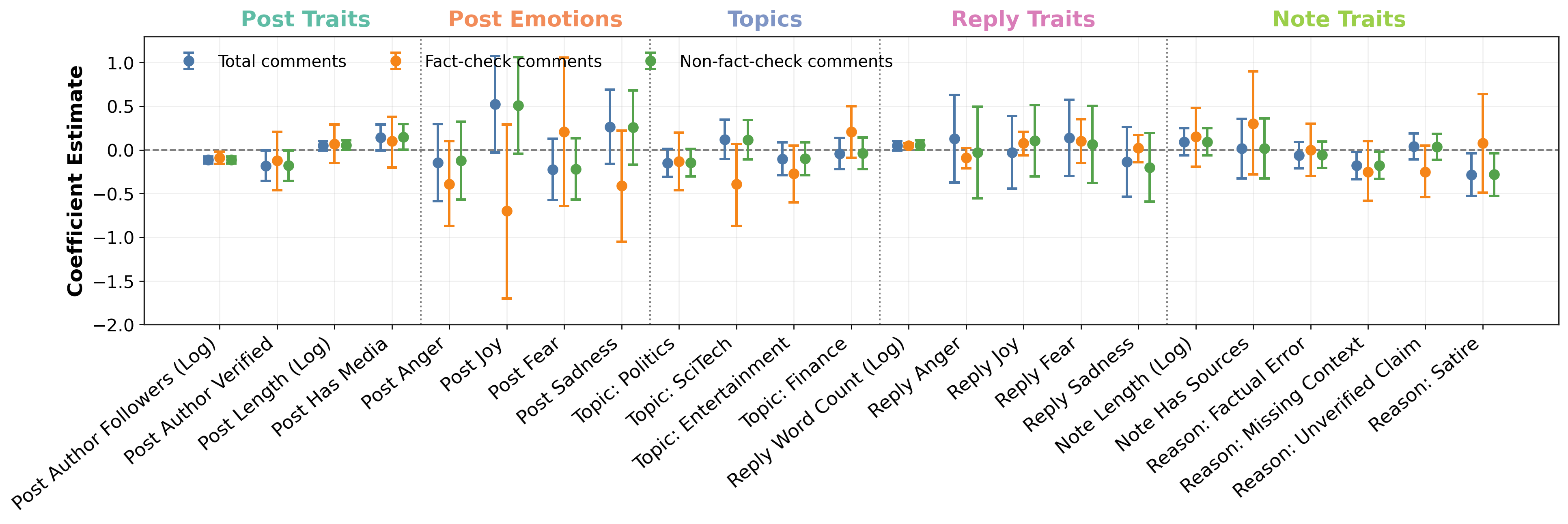}
    \caption{Regression coefficients for the latency of the earliest comment (i.e., FC, non-FC, and all comments) relative to the first community note, analyzed across various factors. Error bars indicate 95\% CIs.}
    \label{fig:result_post_delay}
\end{figure*}

Regarding topical influence, posts about finance uniquely suppress FC comments ($b = -0.374$, $p < .001$) without affecting non-FC comments ($b = -0.071$, $p = .45$), suggesting that financial misinformation presents a higher barrier for users. Political content stimulates a surge in non-FC comments ($b = 0.370$, $p < .001$) but fails to significantly elevate FC comments' volume ($b = 0.203$, $p = .06$). Other topics, such as SciTech ($b = -0.198$, $p =.22$) and Entertainment ($b = 0.235$, $p = .06$), show no significant correlation with FC comments, nor do they increase total comment volume. Similarly, the emotional tone of a post fails to influence FC comments. Anger ($b = -0.372$, $p = .06$), joy ($b = -0.166$, $p = .62$), fear ($b = -0.200$, $p = .30$) and sadness ($b = -0.341$, $p = .08$) lack significant correlations with FC volume. This non-significant pattern mirrors non-FC comments' patterns across anger ($b = -0.425$, $p = .11$), joy ($b = 0.386$, $p = .26$), fear ($b = 0.218$, $p = .27$) and sadness ($b = 0.210$, $p = .33$). 

Community notes' characteristics are not correlated with FC comments' count. Note length ($b = -0.010$, $p = .89$) and the inclusion of trustworthy sources ($b = 0.049$, $p = .65$) show no significant relationship with FC comment volume. In contrast, longer notes significantly reduce non-FC comment volume ($b = -0.243$, $p < .01$), while source inclusion remains non-significant for this group ($b = 0.354$, $p = .05$). These suggest that the perceived complexity to fact-check a post does not influence organic FC comments.

Specific types of misinformation influences FC comments' count. Posts flagged for missing context uniquely attract a higher volume of FC comments ($b = 0.177$, $p < .05$), an effect entirely absent in non-FC comments ($b = 0.073$, $p = .38$). This likely occurs because providing context is more accessible than verifying rigid claims. Conversely, unverified claims ($b = -0.188$, $p < .05$) and satire ($b = -0.302$, $p < .05$) uniquely suppress FC comments without affecting non-FC volume for unverified claims ($b = -0.086$, $p = .33$) or satire ($b = -0.020$, $p = .88$). Factual errors broadly decrease overall engagement by reducing both FC ($b = -0.188$, $p < .01$) and non-FC comments ($b = -0.363$, $p < .001$).

Finally, community note experience-related dimensions suppress comment volume. Both the note author's prior experience ($b = -0.190$, $p < .001$) and the ultimate helpfulness rating of the note ($b = -0.157$, $p < .05$) negatively correlate with FC comment counts. This suppressive effect extends to non-FC comments for both author experience ($b = -0.090$, $p < .001$) and helpfulness ($b = -0.562$, $p < .001$). Experienced authors may produce higher-quality notes that achieve helpful status and public visibility more quickly, thereby resulting in less FC comments.

\textbf{Analysis of FC comments' efficiency}
As shown in Figure~\ref{fig:result_post_delay}, the time delay between the first FC comment and the community note is distinctly robust across multiple post characteristics that otherwise influence general engagement. Whether a post author is verified ($b = -0.126$, $p = .46$), the post length ($b = 0.067$, $p = .56$), and the presence of rich media ($b = 0.096$, $p = .52$) show no significant correlation with FC comments' delay. In contrast, these factors significantly alter the speed of non-FC comments, where verification accelerates responses ($b = -0.193$, $p = .03$), while greater post length ($b = 0.161$, $p < .01$) and media presence ($b = 0.174$, $p < .05$) slow them down. This indicates that the efficiency of FC comments is uniquely robust to post traits. Conversely an author's follower count corresponds to shorter delays for both FC ($b = -0.087$, $p < .05$) and non-FC comments ($b = -0.111$, $p < .001$), suggesting that high visibility accelerates all comments' feedback.

FC comments' speed is also uniquely stable across topical domains. As seen in Figure~\ref{fig:result_post_delay}, Political content significantly accelerates non-FC comments ($b = -0.192$, $p < .05$) but has no significant effect on the timing of FC comments ($b = -0.133$, $p = .44$). Other topics broadly lack significant effects on either FC or general comments' latency, including SciTech (FC $b = -0.397$, $p = .10$; non-FC $b = 0.119$, $p = .30$), Entertainment (FC $b = -0.281$, $p = .09$; non-FC $b = -0.121$, $p = .20$), and Finance (FC $b = 0.210$, $p = .17$; non-FC $b = -0.055$, $p = .55$). Similarly the emotional tone of the source post fails to correlate with the response speed. Anger (FC $b = -0.387$, $p = .12$; non-FC $b = -0.188$, $p = .41$), joy (FC $b = -0.702$, $p = .17$; non-FC $b = 0.441$, $p = .12$), fear (FC $b = 0.210$, $p = .63$; non-FC $b = -0.301$, $p = .10$) and sadness (FC $b = -0.413$, $p = .21$; non-FC $b = 0.217$, $p = .31$) show no influence for both groups.

The attributes of comments also have no correlations with response latency. The word count of a comment does not affect FC comments' ($b = 0.043$, $p = .11$) or non-FC comments' latency ($b = 0.088$, $p = .12$). The emotional intensity within the comments also shows stability across both categories. Anger (FC $b = -0.092$, $p = .11$; non-FC $b = -0.047$, $p = .86$), joy (FC $b = 0.077$, $p = .28$; non-FC $b = 0.074$, $p = .72$), fear (FC $b = 0.100$, $p = .44$; non-FC $b = 0.061$, $p = .78$), and sadness (FC $b = 0.016$, $p = .84$; non-FC $b = -0.225$, $p = .26$) do not alter comments' speed, either.

Finally, FC comments' speed is robust across community-note level factors. Posts flagged for missing context by community notes significantly accelerates non-FC engagement ($b = -0.178$, $p < .05$) but have no such impact on FC latency ($b = -0.246$, $p = .16$). Other note-related factors remain consistently non-significant for both groups, including factual errors (FC $b = 0.002$, $p = .99$; non-FC $b = -0.078$, $p = .31$), unverified claims (FC $b = -0.247$, $p = .11$; non-FC $b = 0.024$, $p = .75$), and satire (FC $b = 0.077$, $p = .79$; non-FC $b = -0.239$, $p = .06$). Basic note attributes similarly do not correlate with response speed, including note length (FC: $b = 0.043$, $p = .11$; non-FC $b = 0.082$, $p = .30$) and the inclusion of trustworthy sources (FC $b = 0.025$, $p = .09$; non-FC $b = 0.024$, $p = .89$).

In summary, we find that: (i) Content richness and emotional tone are associated with FC comments' intensity, and users provide more FC comments in posts ``missing context''. (ii) The speed advantage of organic FC comments is a robust pattern, as opposed to non-FC comments.





\section{Discussions}

Faced with the latency of community notes, our work validates organic comments as rapid corrective signals against misinformation, yielding three main findings: (i) With fine-tuned language models, FC comments can be identified with 91\% accuracy. (ii) FC comments exhibit substantial temporal advantage over community notes. Specifically, 99.4\% of misleading posts receive their first FC comment prior to the creation of a Community Notes, with a median latency of merely 0.1 hours compared to the 9.7-hour delay of community notes. (iii) FC comment intensity uniquely increases with multimedia presence and increasing post length, or when posts are flagged for missing context, while remaining robust across topical and emotional factors. Collectively, our work highlights that proactive user comments can serve as a potential resource for augmenting fact-checking systems.

\textbf{Temporal superiority.} Our findings demonstrate that users' comments provide rapid FC signals that precede the creation of community notes. To ensure fair temporal comparison, we isolate the consensus-building process -- the interval between note writing and display -- and focus on the note creation time~\cite{chuai2024did}. While community notes effectively reduce misinformation spread once displayed, it suffers from delays caused by both note writing and rating~\cite{chuai2024community,chuai2024did}. Prior research indicates that notes typically appear between 15 to 75 hours after a post is created, a critical window during which the majority of viral engagement occurs~\cite{chuai2024community}. Our data reveals the bottleneck even at creation stage, showing a median delay of 9.7 hours for community notes' creation. In contrast, 99.3\% of misleading posts receive their first FC comment within just two hours. This suggests that such ``wisdom of the crowd'' is present during the early stages of a post.

\textbf{Potential of FC comments.} We contribute to the misinformation landscape by establishing the temporal priority of FC comments, providing comparisons between general engagement and FC, and highlighting the proactive nature of FC comments. First, while existing literature on crowdsourced FC emphasizes classification accuracy~\cite{sundriyal2024crowd,shang2020fauxward} or alignment with professional standards~\cite{allen2021scaling}, we prioritize FC comments' temporal availability. By quantifying the robustness of FC comments across topical and emotional factors, those comments provide potential to address the latency gap in the official Community Note program~\cite{chuai2024community}. Second, we highlight the unique FC patterns in comments, expanding beyond general engagement patterns~\cite{solovev2022moral}, or domain-specific findings that fact-checking is slow or absent in COVID-19 misinformation responses~\cite{miyazaki2023fake}. Compared with random posts, we observed significantly more FC comments from misleading posts. Third, our paper could also been seen as a filtering approach to harness the vast volume of law, organic social media comments, potentially complementing those methods that synthesize structured notes~\cite{de2025supernotes}, or identify check-worthy posts through algorithms~\cite{frick2025improved}. Finally, we shift the focus from characterizing behaviors towards using these proactive FC signals. While prior work examined linguistic patterns~\cite{jiang2018linguistic}, debunking behaviors~\cite{miyazaki2023fake}, or backfire effects~\cite{he2024corrective}, we position those proactive comments as prioritization proxies that enable platforms to identify and escalate checkworthy content for review.

\textbf{Vulnerability of FC comments.} However, as our framework primarily detects fact-checking intent, those comments can be vulnerable to false debunking claims or manipulations. For instance, a post describing a violent event might be flooded with fact-like comments claiming that the event is ``fake''. Therefore, we emphasize that organic comments should not be viewed as absolute verification. Instead, they could serve as proactive proxies for prioritization, allowing platforms to identify checkworthy posts for human or algorithmic review. To mitigate the risk of manipulation, these signals could further be integrated with claim normalization~\cite{sundriyal2023chaos}, or consensus mechanisms~\cite{chuai2024did}, before being used in automatic detection models~\cite{nan2024let,nan2025exploiting}.

\textbf{Implications.} Our analysis of FC comments offer three implications. (i) The emergence of comments within hours provides immediate debunking signals, offering a supplementary mechanism for early social correction. Distinct volume and distribution of FC comments also act as proactive, early-warning proxies. Platforms can leverage these engagement patterns to prioritize potentially misleading posts for review, similar to Community Note program's \textit{Request note} function. Furthermore, platforms can incentivize FC comments by highlighting them in reply threads, provided that such visibility does not inadvertently trigger backfire effects~\cite{pilarski2024community,chuai2024did}. (ii) FC comments may augment the Community Note program's recent initiative, such as \textit{Collaborative Notes} or \textit{AI Note Writer}. With a mean of 70 FC comments preceding a visible note, these signals offer a potential resource for automated community note synthesis. Specifically, NLP pipelines can be applied to FC comments to extract recurring counter-claims and specific external URLs. By using a method similar to those synthesizing community notes~\cite{de2025supernotes}, and treating comments as real-time retrieval database, automated models can synthesize note drafts grounded in the crowd response. (iii) FC comments provide preliminary evidence to assist human contributors during their note-drafting process. Looking forward, future frameworks need to enhance reliability by prioritizing comments that explicitly embed external verifiable evidence (e.g., fact-check URLs), using information extraction models to isolate assertions from the FC comments, cross-checking comments using methods similar to the Community Note program's bridging algorithm, or weighing comments based on historical user credibility, analogous to the helpfulness scores of note requestors on $\mathbb{X}$.

\textbf{Limitations.} We acknowledge several limitations of this study. First, regarding selection bias and scope, our analyses focus exclusively on $\mathbb{X}$, and primarily posts receiving community notes up to 2023. This may influence broad generalizability to other social media platforms. Besides, we prioritize direct comments, which capture immediate, organic crowd reactions. This methodology is also consistent with prior research~\cite{chuai2025community}. We recognize that future work should extend this by incorporating nested replies to capture deeper discussions. Finally, our framework targets fact-checking intent rather than veracity. The criteria for classifying FC comments are not precisely aligned with community notes' reasons. The FC comment classification models are also subject to potential bias. To mitigate this, we included checks on the fact-check URLs in comments, which provided additional evidence of the fact-check signals in comments. Future work could integrate automated source-checking (e.g., by \textit{Retrieval-Augmented Generation}) to verify cited evidence, implement dynamic reputation scores to prioritize comments from accounts with a high history of accuracy, and leverage temporal network analysis to flag coordinated inauthentic behavior.

\section{Conclusion}

While community notes provide robust fact-checking, they are slow to prevent the spreading of misleading content. Analyzing over 2.2 million comments, we provide evidence that users' comments contain fact-check intents and appear early. We first constructed an automated framework (93\% accuracy) through fine-tuning language models. We then found that 99.4\% of misleading posts received their first FC comment before the official community note was created. We further found multimedia content triggers a higher volume of FC comments, and this behavior is robust across topical and emotional contexts. These findings suggest that comments serve as a proactive, fast source for augmenting fact-checking on social media.

\bibliography{sample-base}

\clearpage

\section{Ethics Statement}

This study was approved by the Institutional Review Board of [Anonymized]. All analyses used publicly available data and adhered to the respective terms of use. The authors declare no competing interests. We discussed potential ethical and societal concerns below.

\textbf{Human Subjects Protection} For the manual annotation phase, we recruited student assistants. Prior to the task, all participants were explicitly informed of potential risks, specifically the possibility of exposure to toxic, insulting, or offensive content in social media discussions. Participants were guaranteed the right to withdraw from the study at any time without penalty or the need to provide a justification. We strictly maintained participant anonymity by not collecting personally identifiable information (PII). Furthermore, before the annotation, we clearly informed participants that they could request the deletion of their annotation data upon withdrawal or at any subsequent time.

\textbf{Data Protection} For the experiment data and annotation data, we stored them in a dedicated and segregated server, without sharing the data with people outside the author group. The annotation website is only disseminated to these student assistants, and the website was disabled promptly after the annotation task. The server enforced authentication to access.

\textbf{Data and Model Licensing} Our utilization of the X community notes dataset strictly adhered to the X Developer Agreement\footnote{https://developer.x.com/en/developer-terms/agreement-and-policy}. Furthermore, all pre-trained models were employed in compliance with their respective open-source licenses: BERT and SciBERT are governed by the Apache 2.0 License, while RoBERTa, DeBERTa, and E5 follow the MIT License.

\textbf{Societal Impact} While our framework aims to mitigate misinformation, we acknowledge potential negative societal impacts. First, our detection relies on linguistic patterns of reasoning rather than the factual veracity of the comment. Consequently, there is a risk that confidently phrased but inaccurate user comments could be misidentified as valid corrections, potentially amplifying misinformation if deployed without human oversight. Second, public knowledge of these detection features could enable malicious actors to craft adversarial content that mimics fact-checking styles to evade moderation or manipulate community consensus.

\section{Significance Table of Fact-Checking and Non-Fact-Checking Comments}

Table~\ref{tab:count_model} and ~\ref{tab:latency_model} showed the findings across the count model and the latency model.

\begin{table*}[htbp]
\centering
\caption{The coefficients of the count regression model.}
\label{tab:count_model}
\resizebox{\textwidth}{!}{%
\begin{tabular}{ll c c c c}
\toprule
\textbf{Variable} & \textbf{Significance} & \textbf{FC $b$ [95\% CI]} & \textbf{FC $p$-value} & \textbf{Non-FC $b$ [95\% CI]} & \textbf{Non-FC $p$-value} \\ 
\midrule

\multicolumn{6}{l}{\textit{\textbf{Post Traits}}} \\
\midrule
Post Author Followers (Log) & \both & 0.212 [0.163, 0.261] & \psig{$<$0.001} & 0.228 [0.188, 0.268] & \psig{$<$0.001} \\
Post Author Verified        & \neither & -0.199 [-0.424, 0.025] & 0.082 & 0.004 [-0.191, 0.199] & 0.969 \\
Post Length (Log)           & \fconly & 0.255 [0.111, 0.398] & \psig{$<$0.001} & 0.062 [-0.070, 0.194] & 0.358 \\
Post Has Media              & \fconly & 0.350 [0.146, 0.554] & \psig{$<$0.001} & -0.051 [-0.218, 0.116] & 0.550 \\
\midrule

\multicolumn{6}{l}{\textit{\textbf{Post Emotions}}} \\
\midrule
Post Anger   & \neither & -0.372 [-0.766, 0.022] & 0.064 & -0.425 [-0.938, 0.088] & 0.105 \\
Post Joy     & \neither & -0.166 [-0.828, 0.496] & 0.623 & 0.386 [-0.279, 1.051] & 0.256 \\
Post Fear    & \neither & -0.200 [-0.575, 0.175] & 0.296 & 0.218 [-0.173, 0.608] & 0.274 \\
Post Sadness & \neither & -0.341 [-0.727, 0.044] & 0.083 & 0.210 [-0.212, 0.632] & 0.329 \\
\midrule

\multicolumn{6}{l}{\textit{\textbf{Topics}}} \\
\midrule
Topic: Politics      & \nfconly & 0.203 [-0.011, 0.417] & 0.063 & 0.370 [0.192, 0.548] & \psig{$<$0.001} \\
Topic: SciTech       & \neither & -0.198 [-0.515, 0.118] & 0.220 & 0.010 [-0.273, 0.292] & 0.947 \\
Topic: Entertainment & \neither & 0.235 [-0.012, 0.482] & 0.062 & 0.097 [-0.109, 0.302] & 0.357 \\
Topic: Finance       & \fconly & -0.374 [-0.605, -0.143] & \psig{$<$0.001} & -0.071 [-0.256, 0.115] & 0.454 \\
\midrule

\multicolumn{6}{l}{\textit{\textbf{Note Traits}}} \\
\midrule
Note Length (Log)        & \nfconly & -0.010 [-0.144, 0.125] & 0.885 & -0.243 [-0.413, -0.072] & \psig{0.005} \\
Note Has Sources         & \neither & 0.049 [-0.162, 0.260] & 0.649 & 0.354 [-0.001, 0.709] & 0.051 \\
Reason: Factual Error    & \both & -0.188 [-0.324, -0.053] & \psig{0.006} & -0.363 [-0.534, -0.192] & \psig{$<$0.001} \\
Reason: Missing Context  & \fconly & 0.177 [0.041, 0.313] & \psig{0.011} & 0.073 [-0.088, 0.233] & 0.377 \\
Reason: Unverified Claim & \fconly & -0.188 [-0.340, -0.035] & \psig{0.016} & -0.086 [-0.259, 0.087] & 0.329 \\
Reason: Satire           & \fconly & -0.302 [-0.537, -0.066] & \psig{0.012} & -0.020 [-0.282, 0.243] & 0.884 \\
Author Exp (Log)         & \both & -0.190 [-0.252, -0.128] & \psig{$<$0.001} & -0.090 [-0.129, -0.051] & \psig{$<$0.001} \\
Note Is Helpful          & \both & -0.157 [-0.284, -0.030] & \psig{0.015} & -0.562 [-0.757, -0.367] & \psig{$<$0.001} \\
\bottomrule
\end{tabular}%
}
\end{table*}

\begin{table*}[htbp]
\centering
\caption{The coefficients of the latency regression model.}
\label{tab:latency_model}
\resizebox{\textwidth}{!}{%
\begin{tabular}{ll c c c c}
\toprule
\textbf{Variable} & \textbf{Significance} & \textbf{FC $b$ [95\% CI]} & \textbf{FC $p$-value} & \textbf{Non-FC $b$ [95\% CI]} & \textbf{Non-FC $p$-value} \\ 
\midrule

\multicolumn{6}{l}{\textit{\textbf{Post Traits}}} \\
\midrule
Post Author Followers (Log) & \both & -0.087 [-0.161, -0.013] & \psig{0.021} & -0.111 [-0.151, -0.071] & \psig{$<$0.001} \\
Post Author Verified        & \nfconly & -0.126 [-0.462, 0.210] & 0.462 & -0.193 [-0.368, -0.019] & \psig{0.030} \\
Post Length (Log)           & \nfconly & 0.067 [-0.155, 0.289] & 0.555 & 0.161 [0.049, 0.272] & \psig{0.005} \\
Post Has Media              & \nfconly & 0.096 [-0.197, 0.388] & 0.520 & 0.174 [0.024, 0.325] & \psig{0.023} \\
\midrule

\multicolumn{6}{l}{\textit{\textbf{Post Emotions}}} \\
\midrule
Post Anger   & \neither & -0.387 [-0.870, 0.095] & 0.116 & -0.188 [-0.633, 0.256] & 0.406 \\
Post Joy     & \neither & -0.702 [-1.695, 0.291] & 0.166 & 0.441 [-0.107, 0.988] & 0.115 \\
Post Fear    & \neither & 0.210 [-0.640, 1.060] & 0.628 & -0.301 [-0.657, 0.054] & 0.097 \\
Post Sadness & \neither & -0.413 [-1.050, 0.225] & 0.205 & 0.217 [-0.202, 0.636] & 0.310 \\
\midrule

\multicolumn{6}{l}{\textit{\textbf{Topics}}} \\
\midrule
Topic: Politics      & \nfconly & -0.133 [-0.466, 0.201] & 0.435 & -0.192 [-0.356, -0.029] & \psig{0.021} \\
Topic: SciTech       & \neither & -0.397 [-0.864, 0.070] & 0.096 & 0.119 [-0.106, 0.343] & 0.301 \\
Topic: Entertainment & \neither & -0.281 [-0.608, 0.046] & 0.092 & -0.121 [-0.309, 0.066] & 0.204 \\
Topic: Finance       & \neither & 0.210 [-0.087, 0.506] & 0.165 & -0.055 [-0.235, 0.125] & 0.548 \\
\midrule

\multicolumn{6}{l}{\textit{\textbf{Reply Traits}}} \\
\midrule
Reply Word Count (Log) & \fconly & 0.043 [-0.010, 0.095] & 0.108 & 0.088 [-0.023, 0.200] & 0.121 \\
Reply Anger            & \neither & -0.092 [-0.205, 0.020] & 0.109 & -0.047 [-0.569, 0.475] & 0.860 \\
Reply Joy              & \neither & 0.077 [-0.062, 0.217] & 0.279 & 0.074 [-0.328, 0.477] & 0.718 \\
Reply Fear             & \neither & 0.100 [-0.152, 0.352] & 0.437 & 0.061 [-0.373, 0.496] & 0.783 \\
Reply Sadness          & \neither & 0.016 [-0.139, 0.170] & 0.839 & -0.225 [-0.617, 0.168] & 0.262 \\
\midrule

\multicolumn{6}{l}{\textit{\textbf{Note Traits}}} \\
\midrule
Note Length (Log)        & \neither & 0.043 [-0.010, 0.095] & 0.108 & 0.082 [-0.074, 0.238] & 0.304 \\
Note Has Sources         & \neither & 0.025 [-0.004, 0.053] & 0.086 & 0.024 [-0.319, 0.367] & 0.890 \\
Reason: Factual Error    & \neither & 0.002 [-0.307, 0.310] & 0.990 & -0.078 [-0.229, 0.072] & 0.305 \\
Reason: Missing Context  & \nfconly & -0.246 [-0.589, 0.096] & 0.159 & -0.178 [-0.334, -0.023] & \psig{0.025} \\
Reason: Unverified Claim & \neither & -0.247 [-0.546, 0.052] & 0.105 & 0.024 [-0.125, 0.174] & 0.750 \\
Reason: Satire           & \neither & 0.077 [-0.489, 0.643] & 0.790 & -0.239 [-0.486, 0.008] & 0.057 \\
\bottomrule
\end{tabular}%
}
\end{table*}

\end{document}